\documentclass[draftclsnofoot,onecolumn]{IEEEtran}
\usepackage{amsmath,amsfonts}
\usepackage{algorithmic}
\usepackage{booktabs}
\usepackage{tabularx}
\usepackage{algorithm}
\usepackage{array}
\usepackage[caption=false,font=normalsize,labelfont=sf,textfont=sf]{subfig}
\usepackage{textcomp}
\usepackage{stfloats}
\usepackage{url}
\usepackage{verbatim}
\usepackage{graphicx}
\usepackage{cite}

\begin{document}

\title{\huge Approaching Shannon's One-Time Pad: Metrics, Architectures, and Enabling Technologies}

\author{Li Sun, Xianhui Lu, Peng Liu, Jianjun Wu, Xiaohu You, and Xiaofeng Tao
	\thanks{This work was supported in part by the National Key R\&D Program under Grant 2020YFB1807502. This work has been submitted to the IEEE for possible publication. Copyright may be transferred without notice, after which this version may no longer be accessible.}
	\thanks{Li Sun, Peng Liu, and Jianjun Wu are with Wireless Technology Lab, 2012 Labs, Huawei Technologies, Shenzhen 518000, China. \emph{Corresponding author:} Li Sun (sunli50@huawei.com)}
	\thanks{Xianhui Lu is with the Institute of Information Engineering, Chinese Academy of Sciences, Beijing 100085, China.}
	\thanks{Xiaohu You is with the the National Mobile Communications Research Laboratory, Southeast University, Nanjing 211189, China; and also with Purple Mountain Laboratories, Nanjing 211189, China.}
	\thanks{Xiaofeng Tao is with the School of Information and Communication Engineering, Beijing University of Posts and Telecommunications, Beijing 100876, China; and also with the Department of Broadband Communication, Peng Cheng Laboratory, Shenzhen 518055, China.}
}

\markboth{Submitted to XXX}%
{Sun \MakeLowercase{\textit{et al.}}: Approaching Shannon's Perfect Secrecy}


\maketitle

\begin{abstract}
The rapid development of advanced computing technologies such as quantum computing imposes new challenges to current wireless security mechanism which is based on cryptographic approaches. To deal with various attacks and realize long-lasting security, we are in urgent need of disruptive security solutions. In this article, novel security transmission paradigms are proposed to approach Shannon's one-time pad perfect secrecy. First, two metrics, termed as Degree-of-Approaching (DoA) and Degree-of-Synchronous-Approaching (DoSA), are developed to characterize the closeness between the achieved security strength and perfect secrecy. These two metrics also serve as a guideline for secure transmission protocol design. After that, we present two paths towards Shannon's one-time pad, i.e., an explicit-encryption based approach and an implicit-encryption based approach. For both of them, we discuss the architecture design, enabling technologies, as well as preliminary performance evaluation results. The techniques presented in this article provide promising security-enhancing solutions for future wireless networks.
\end{abstract}


\section{Introduction}
Security is always one of the major concerns for wireless network design. Current network security solutions highly rely on cryptographic approaches deployed at upper layers of the protocol stack. While being adopted widely in almost all generations of mobile communications systems, this paradigm might not be competent for guaranteeing communications confidentiality in future wireless networks such as the sixth-generation (6G). The reasons are twofold.

First, modern cryptography is built upon the complexity theory, and the unbreakability of the existing encryption algorithms is dependent of the assumption that the attacker has limited computation capability. To be more specific, \emph{given polynomially bounded computational resources}, on average it is hard for an adversary to extract any information about the plaintext from the ciphertext. With the development of computing technologies (especially the fast evolution of quantum computing), the security of classical encryption methods is being challenged. For example, to break the Rivest-Shamir-Adleman (RSA) algorithm, sub-exponential-time computation is required for electronic computers, but only polynomial-time computation is needed if a quantum computer is used. While rapid progress has been made in developing post-quantum cryptographic algorithms, the security of these algorithms essentially arises from the difficulty in solving hard problems in mathematics. The potential risks of this security mechanism become increasingly apparent with the progress in mathematical theories and the proposal of advanced algorithms. This is exemplified by the recent event that the post-quantum cryptographic algorithm SIDH was cracked within only an hour using a single-core CPU \cite{Castryck}. Therefore, it is urgent to develop disruptive techniques to cope with quantum attacks and provide long lasting security.

Second, due to the openness of wireless medium, radio access network (RAN) is vulnerable to various attacks \cite{Ahmad}. In current systems, almost all the security protections are deployed at higher layers (e.g., the Packet Data Convergence Protocol (PDCP) layer), and there is a lack of encryption or integrity protection at MAC and PHY layers. As a consequence, various messages (e.g., control signals) generated by MAC or PHY layers are subject to eavesdropping, jamming, or modifications, yielding the malfunctions of network entities. While in theory higher-layer encryption algorithms can be applied to solve this problem, additional overhead and processing delay are prohibited in many applications such as industrial control, vehicular communications, Internet-of-Things (IoT), etc.


As another secure communications mechanism, physical layer security (PLS) has been intensively studied by the academia in the past years \cite{Poor}-\cite{Sun}. Different from the traditional complexity-based security paradigm, PLS exploits the intrinsic characteristics of wireless channels, such as random noise, fading, and interference, to degrade the received signal quality at the malicious users, and realizes secure transmission via signal design and signal processing approaches. While fruitful achievements have been made so far in PLS research, there is still lack of rigorous and practical methodologies to evaluate the security performance of PLS solutions. The widely-adopted performance metrics in PLS can be divided into two categories. The first category is the information-theoretic metrics, including secrecy capacity, secrecy outage probability, etc. These metrics only characterize the theoretical limits of secure transmissions, which cannot be applied to performance evaluation of practical systems. Another popular metric is bit error rate (BER) or symbol error rate (SER). With these metrics, the achievable security strength can be measured by the level of the error floor at the eavesdropper. Although this metric is of practical significance, its rigorousness is always criticized by security experts, because it is unclear how much information is leaked given the achievable error rate at the eavesdropper.

To deal with the challenges mentioned above, enhanced security mechanisms with provable security performance are required to be devised. If looking back to the history of security technologies, we can find that the modern security technologies originate from Shannon's seminal paper \emph{Communication Theory of Secrecy Systems}, published in 1949 \cite{Shannon}. In this pioneering work, Shannon gave the definition of perfect secrecy from an information-theoretic perspective. Consider a system consisting of a pair of legitimate transceivers Alice and Bob and an eavesdropper Eve. According to Shannon's definition, this system is said to be in perfect secrecy if the following condition is satisfied:
\begin{equation}
\label{eq-1}
H(M|X)=H(M),
\end{equation}
where $H(M)$ and $H(M|X)$ represent the entropy of Alice's transmitted message \emph{M} and the conditional entropy of \emph{M} given Eve's observation \emph{X}, respectively. Intuitively, (1) indicates that Eve's uncertainty about \emph{M} is not decreased after it intercepts the legitimate user's signal, which means that the amount of leaked information is zero. Perfect secrecy can be achieved by a simple procedure called one-time pad (OTP). In OTP, a secret key is first pre-shared between Alice and Bob securely, i.e., the secret key is only known by the legitimate users but not accessible by Eve. Then, each message bit is XORed with a separate key bit to produce the corresponding encrypted bit. Mathematically, this operation can be described as $X=M\oplus K$. It has been shown that as long as the key bits are independent and uniformly distributed, the codeword \emph{X} is statistically independent of the message \emph{M}, which naturally guarantees that $H(M|X)=H(M)$.

It is interesting to compare the OTP approach with the classical cryptographic alternatives, which follow totally different design philosophies. For OTP, both encryption and decryption are realized using a very simple algorithm (i.e., XOR), and the security of the OTP method totally relies on the freshness of the secret key (Each key can be used only once!). Comparably, in classical cryptography-based solutions, the key does not change for a long time, and the security is guaranteed by highly complicated encryption/decryption algorithms. OTP is a perfect-secrecy achieving strategy and is thus able to offer long lasting security. However, the cost of implementing OTP is prohibitively high because the legitimate parties must generate, store, and securely share long keys consisting of random bits. As opposed to this paradigm, the cryptography based approaches are implementable in practice, which has been evidenced by the successful deployment of cryptographic protocols in all generations of mobile communications systems. Nevertheless, security risks faced by these approaches become increasingly apparent with the emergence of advanced computing technologies. It is thus demanding to combine cryptography theory and PLS techniques to develop novel security techniques that can ``approach" OTP perfect secrecy while maintaining an affordable cost. This motivates the work in this article.

The rest of this paper is organized as follows. In Sect. II, the concept of ``approaching one-time pad" is presented, which gives performance metrics for evaluating novel security solutions. Sect. III and Sect. IV proposed two paths towards perfect secrecy, namely an explicit-encryption based method and an implicit-encryption based method. The comparison of the existing secure communications techniques and the proposed solutions is given in Sect. V. Finally, we conclude this paper in Sect. VI.

\section{Approaching OTP: The Metrics}
To develop novel security solutions to approach OTP, we must first clarify what ``approaching OTP" means. For any security-provisioning technique, we introduce the terminology ``Degree-of-Approaching (DoA)" to measure its achieved security strength, i.e., the closeness of its security level to perfect secrecy. The DoA is defined as
\begin{equation}
\label{eq-2}
D=\frac{E_{K}}{E_{M}},
\end{equation}
where $E_{K}$ and $E_{M}$ denote the entropy of the key and that of the information message to be protected, respectively. DoA characterizes the achieved security level of a communications system. To be specific, the larger the DoA is, the higher the security level will be. On one hand, given $E_{M}$ (i.e., the total amount of information that needs to be protected), larger values of \emph{D} imply longer keys and thus increased difficulty in deciphering the encrypted message. On the other hand, for a fixed $E_{K}$, the total number of information bits that is protected by these $E_{K}$ key bits (i.e., $E_{M}$) is inversely proportional to \emph{D}, which suggests that the amount of leaked information due to key exposure is small (large) for large (small) values of \emph{D}. For most practical communications systems where classical cryptographic methods are employed, the achieved DoA is rather low (e.g., close to zero) because the root key used to generate key streams do not change over a long period. In comparison, the DoA of the OTP approach is 1 because the amount of key bits is the same as that of information bits.

It should be emphasized that the aforementioned definition is general. First, the so-called ``key" in this definition may have various forms. It can be distributed by upper layers of the protocol stack (as is done in current commercial communications systems), or extracted from wireless channels and (or) hardware. Even, it can also be ``virtual keys" that are generated from the bit errors created at the eavesdropper by PLS techniques, which will be elaborated on later. The encryption algorithm mentioned here can be either the existing symmetric encryption algorithms such as Advanced Encryption Standard (AES), the OTP approach, or the implicit encryption method which combines PLS with randomness extraction algorithm to offer provable security strength (discussed later).

While the DoA can characterize the security strength, it does not describe the efficiency in approaching perfect secrecy. Consider a specific example where a communications system delivers 1Mbits information within 1 second. To achieve a DoA of 1, 1M secret key bits are required to be generated. However, it might take an hour to produce such a huge number of key bits. This example indicates that a high DoA may be realized with a significant degradation in communications efficiency. It is not difficult to achieve a DoA of 1 if the cost is not taken into account. In this sense, the achieved DoA does not reflect the intrinsic security-provisioning capability of a communications system. Instead, it is only determined by specific considerations in system design. This motivates us to introduce a new metric which takes into account not only the security level but also the ``approaching efficiency", called Degree-of-Synchronous-Approaching (DoSA).

The DoSA is defined as
\begin{equation}
\label{eq-3}
d=\frac{C_{K}}{C_{M}},
\end{equation}
where $C_{K}$ and $C_{M}$ are the secret-key capacity (i.e., the maximum amount of secret-key bits extracted per unit time) and the channel capacity (i.e., the maximum amount of information bits transmitted per unit time), respectively.

Essentially, DoSA measures the system's intrinsic capability in approaching OTP perfect secrecy. Given the channel capacity (i.e., the achievable information transmission rate), the larger DoSA is, the higher the key generation rate will be. Note that a high rate in key generation actually manifests that the system is able to offer strong security protections for high data-rate services. From another perspective, DoSA can also be used as an indication of system's efficiency in approaching the OTP perfect secrecy. For a pre-determined target DoA corresponding to a requirement on security strength, a large value in DoSA implies a high ratio in information transmission time to key generation time. Since the incurred overhead for realizing the target security level is inversely proportional to the aforementioned ratio, DoSA can be viewed as a metric for assessing the OTP approaching efficiency.

By utilizing the metrics of DoA and DoSA, the achievable security strength of a communications system can be evaluated more precisely. Moreover, guided by these new metrics, novel security paradigm can be devised beyond the existing cryptography based solutions. As is shown from Fig. 1, strong secrecy, meaning that the DoA is between 0 and 1, can be realized by combining the existing symmetric encryption algorithms with frequently-updated keys (e.g., the system may operate in a ``one key per packet" mode). The key update period and (or) the ratio of key bits to information bits per packet can be adjusted to balance the achieved security level and the implementation overhead. Compared to the two ``extreme" solutions, i.e., OTP and classical cryptographic approaches with DoAs being 1 and close to zero respectively, the ``Approaching OTP" solution enables a more flexible security configuration, and is more adaptive to users' diversified demands in security, communication rate, and cost.

\begin{figure}[!t]
\centering
\includegraphics[width=5in]{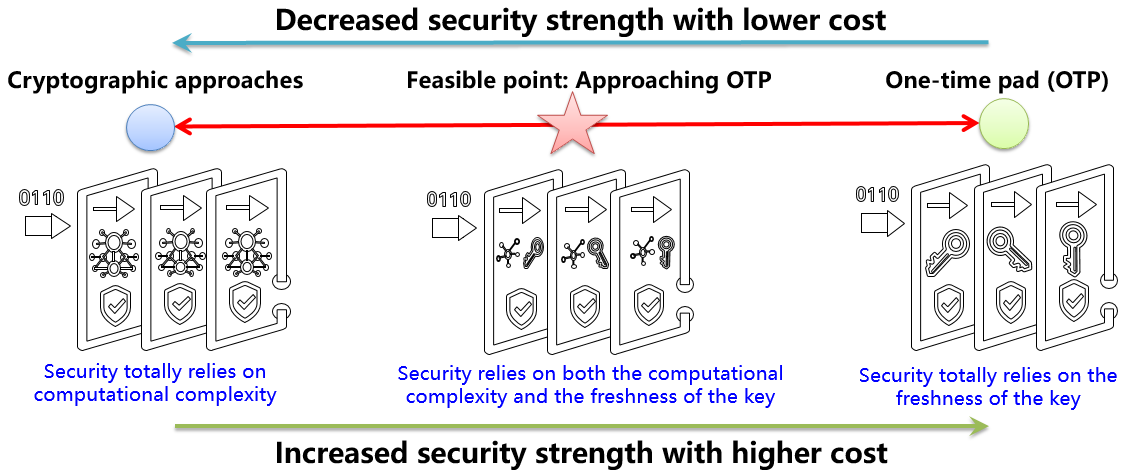}
\caption{Illustration of Approaching-OTP.}
\label{fig_1}
\end{figure}

\section{The Method of Approaching OTP: Explicit Encryption Based Approach}
A natural solution of approaching OTP is based on explicit encryption, where secret keys are first derived from wireless propagation environment and (or) transceiver hardware, and then encryption is realized using one-time pad or other symmetric encryption algorithms. While this idea is natural, one critical technical issue needs to be addressed: How to generate a large number of random key bits within a short period such that the key generation rate well matches the information transmission rate?

\subsection{Architecture Design for High-Rate Key Generation}

Physical layer key generation has been studied by the academia for several decades\cite{Zeng}-\cite{KeyG}. The mainstream framework for key generation is based on channel reciprocity, for which the core idea is that legitimate communications parties separately derive secret key bits from their observations of the channel. Under the ideal assumption that the channel reciprocity holds and the channel estimation is perfect, in theory the two communications devices can generate identical key bits. As demonstrated by the left part of Fig. 2, channel-reciprocity based key generation (CBKG) procedure consists of four steps, namely channel measurement, quantization, information reconciliation, and privacy amplification. In the first step, the communication parties measure the radio channel by sending pilot signals. After that, the measured channel coefficients are quantized into binary sequences. Due to inevitable measurement errors and channel noises, the generated binary sequences at the two sides might slightly differ from each other, which disables direct use of these bits as secret keys. To solve this problem, information reconciliation is then activated, where the two communications parties correct inconsistent information bits through negotiation to generate consistent key sequences. Finally, privacy amplification is employed to remove the revealed information from the agreed key sequence at the communication pairs. There are many methods implementing the privacy amplification, such as using Hash functions.

\begin{figure}[!t]
\centering
\includegraphics[width=4.5in]{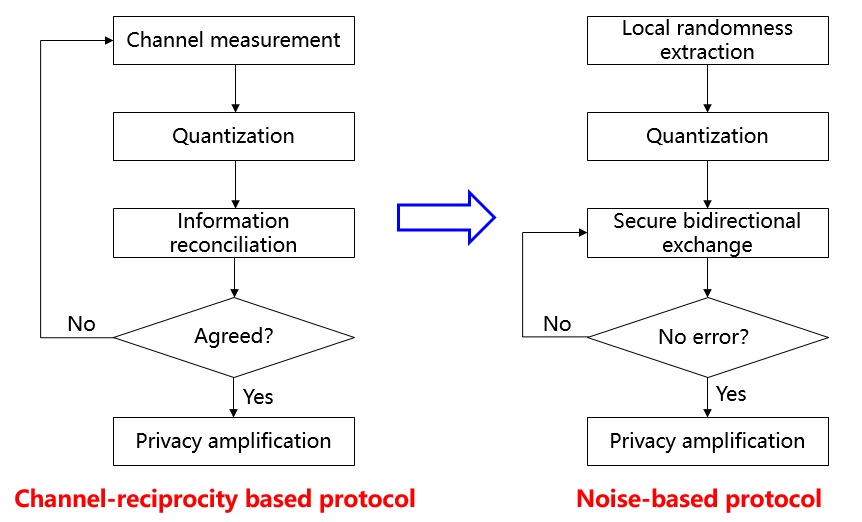}
\caption{Channel-reciprocity based key generation architecture versus noise-based key generation architecture.}
\label{fig_2}
\end{figure}

The channel-reciprocity based approach mentioned above has attracted much research attention, and the majority of the physical layer key generation algorithms are based on this framework. However, as has been pointed out by multiple articles, the major drawback of this classical framework is that the key generation rate is too low \cite{Zhang}. To be more specific, because of the high correlation in time, frequency, and spatial domains of the wireless channel, it is rather difficult to distill enough random bits from channels within a short time period. This yields that the key generation rate is much lower than the information transmission rate, resulting in an extremely low DoSA. To combat this barrier, a noise-based key generation (NBKG) architecture that does not rely on channel properties is devised, the details of which are shown below.

Instead of using channel coefficients as the ``materials" to produce key bits, in NBKG, natural noises are used as the random source for key generation. More specifically, noise samples, coming from either the hardware circuits or the environments, are first individually collected by two communications parties (called Alice and Bob hereafter). Based on these noise samples, corresponding local keys are separately derived using quantization algorithms. Then, the obtained local keys are shared between Alice and Bob to produce the global key sequence, which will be used for future encryption. During the local-key sharing procedure, physical layer security approaches are adopted to guarantee that the eavesdropper is unable to derive the key based on its observations, the details of which will be given shortly. The NBKG protocol is comprised of four steps as illustrated in the right part of Fig. 2, i.e., local randomness extraction, quantization, secure bidirectional exchange of local key, and privacy amplification. Here, the purposes of quantization and privacy amplification are the same as those in the classical reciprocity-based framework.

The differences between the NBKG architecture and the CBKG alternative lie in Step 1 and Step 3. For Step 1, various random sources, e.g., the out-of-band received signal that contains noise samples, or the thermal noises of the hardware, can be exploited to produce the local key, and we will not discuss much about this issue. The critical challenge in implementing NBKG lies in Step 3. To prevent the eavesdropper from intercepting the exchanged local keys, a constellation-rotation aided approach is devised, which will be elaborated on in what follows.

\subsection{Secure Bidirectional Exchange of Local Keys}
Here we consider a full-duplex communications system where both end devices can transmit and receive simultaneously over the same bandwidth. First, each device encodes its local key bits and modulates the coded bits into complex symbols. Afterwards, the complex constellation point is rotated with an appropriate angle, and is then projected onto the real axis to obtain a real point. As long as the rotation angle is properly chosen, the obtained real signal can fully represent the original complex version as illustrated by Fig. 3, which implies that it suffices to transmit these projections instead of the original complex constellation points. In this manner, an additional transmission degree-of-freedom is created, which can be used for secrecy enhancement. Finally, the transmitted complex signal is generated. Its real component is the aforementioned projection, and the imaginary component is left for injecting the artificial noise, which is used to degrade the received signal quality at the eavesdropper and secure the key exchange procedure. At Alice (Bob), after self-interference cancellation is completed, the information-bearing signal component that carries the local key lies orthogonally to the artificial noise component sent from Bob (Alice), thus not degrading the signal detection performance of the legitimate link. On the other hand, at the eavesdropper, thanks to the signal superposition effect created by full-duplex transmission, the artificial noise component and the information-bearing component are mixed together in the 2D signal space, thereby yielding a high error floor and significantly worsening Eve's decoding performance. This can be clearly seen from the received constellations at different nodes in Fig. 3.

\begin{figure}[!t]
\centering
\includegraphics[width=4.5in]{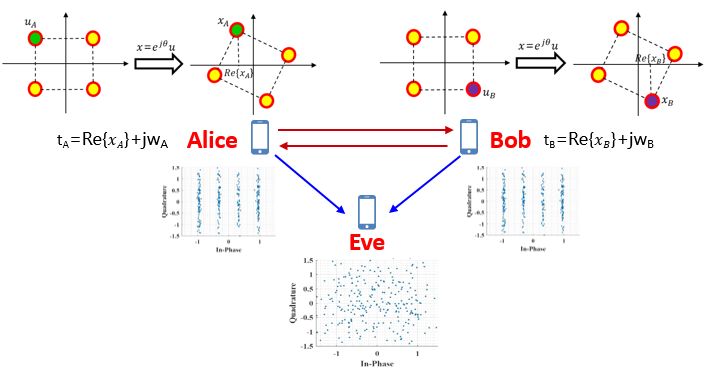}
\caption{Constellation-rotation aided secure transmission method.}
\label{fig_3}
\end{figure}

\textbf{\emph{Remark: }} Careful readers may have the following question: Now that the proposed method can be used to securely exchange local keys, why not directly using this method to realize secure bidirectional message transmissions? Here we would like to emphasize that while in theory the proposed method can be utilized to secure message transmissions, it is not suggested to do so because of two reasons. First, the use of constellation rotation and artificial noise injection techniques incurs non-negligible rate penalty for the legitimate link, which is not acceptable for most practical applications. However, for key-exchange, rate loss is not a significant issue. Second, although a high error floor can be created at the eavesdropper using the proposed scheme, the eavesdropper is still able to obtain a small amount of the exchanged bits, which results in information leakage if this method is used for data transmissions. Nevertheless, when being applied to key generation, thanks to the privacy amplification operation (e.g., using Hash functions) in the final step of the protocol, the bit errors introduced at the eavesdropper will cause the global key generated at the eavesdropper to be totally different from that at the legitimate users. In other words, the proposed algorithm is competent for preventing the key from being intercepted, but is not sufficient to guarantee message confidentiality.

\subsection{Performance Evaluation}
Based on published results in the literature and the definition of the DoSA given in Sect. II, it can be straightforwardly obtained that the DoSA of the CBKG approaches is typically close to zero for most practical systems. Compared to this, by using the constellation-rotation method, in high SNR regime the achievable DoSA of the proposed NBKG framework can be approximated as \footnote{For ease of derivations, it is assumed that the self-interference cancellation is perfect, the channel coefficients of all legitimate links are 1, and the noise variances at all nodes are the same.}
\begin{eqnarray}
\label{eq-4}
d=\frac{I(X_{K,1};Y_{K,2})+I(X_{K,2};Y_{K,1})-I((X_{K,1},X_{K,2});Z)}{I(X_{1};Y_{2})+I(X_{2};Y_{1})}\approx  1-\frac{2.83}{\log \mathrm{SNR}},
\end{eqnarray}
where $\mathrm{SNR}$ is the ratio of the transmit power to the noise variance at the receiver. It is clear that the achievable DoSA approaches 1 for high SNRs, which fully demonstrates that the proposed NBKG framework is able to provide security guarantee for high-rate data transmissions.

To evaluate the performance of the proposed techniques, a demo system is developed, as is shown in the upper part of Fig. 4. The experimental results for image transmission are displayed in the lower left and lower right parts of Fig. 4, which exhibit the results at the legitimate user and the eavesdropper, respectively. We observe from Fig. 4 that by using the proposed techniques, no decoding error is incurred at the legitimate devices, and the transmitted image can be recovered perfectly. In comparison, the constellation points are fully obfuscated at the eavesdropper, preventing the eavesdropper from recovering any useful information of the image.

\begin{figure}[!t]
\centering
\includegraphics[width=4.5in]{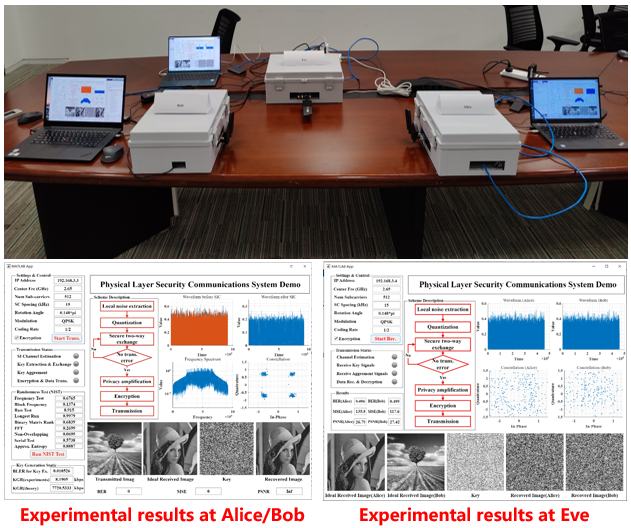}
\caption{Experimental results of the proposed secure transmission scheme.}
\label{fig_4}
\end{figure}

\subsection{Comparisons Between Two Architectures}

\begin{table}[ht]
\begin{center}
\caption{\label{tab:table1}Comparison of two key generation architectures}
\begin{tabular}{ccc}
\toprule
Performance metrics & CBKG & NBKG \\
\midrule
Randomness & Low & High\\
Key generation rate & Low &High\\
Key mismatch rate & No guarantee & Guarateed\\
Anti-eavesdropping ability & Weak & Strong\\
\bottomrule
\end{tabular}
\end{center}
\end{table}

As a summary, we compare the proposed NBKG architecture with the classical CBKG alternative from four aspects. First, NBKG uses intrinsic noises extracted from devices or communications medium as the materials of the secret key, which guarantees that the generated key sequence is truly random. In comparison, it is much more difficult for the CBKG methods to produce key sequence with high randomness, which is attributed to the strong correlations of the wireless channels in time, frequency, and spatial domains. Second, the key generation rate achieved by the proposed method is only constrained by the information transmission rate (In practice, the key generation rate is slightly lower than the information transmission rate, due to the rate penalty caused by the use of physical layer security transmission algorithms.), which can be as high as several Mbps or above for modern communications systems. In contrast, the key generation rate of the CBKG approaches is much lower than the information transmission rate \cite{Zhang}. Third, the NBKG method has a guaranteed key mismatch rate, which is roughly equal to the transmission error rate. Comparably, there is no guarantee on the key mismatch rate for the CBKG approach. In particular, the generated key bits at two sides cannot be identical to each other without information reconciliation. Even when information reconciliation is employed, it is still not sure whether or not the keys derived by the two parties are the same. Fourth, in terms of the anti-eavesdropping capability in the key generation procedure, the proposed method is much stronger thanks to the use of physical layer security techniques. In comparison, the CBKG protocol is more vulnerable to eavesdropping because the wiretappers located nearby the legitimate nodes may have similar channel observations compared to the legitimate user, from which similar key bit sequence can be easily derived. The comparison between the proposed NBKG protocol and the classical CBKG protocol is briefly summarized in Table I.

\section{The Method of Approaching OTP: Implicit Encryption Based Approach}
Different from the explicit encryption paradigm, implicit encryption does not rely on key generation or encryption/decryption operations. Instead, it takes advantages of the features of wireless channels to create a high error floor, or equivalently introduce randomness, at the eavesdropper, thus preventing the eavesdropper from decoding the source message.

It appears that all of the existing PLS transmission techniques can be viewed as implicit encryption methods. However, this is not true. As has been discussed in Section I, due to the absence of appropriate metrics, PLS schemes developed so far primarily follow heuristic design methodologies, and are unable to provide \emph{strictly provable} security levels. Therefore, when developing implicit-encryption based secure transmission framework, a new transmission architecture should be devised which realizes provable security without using keys. Moreover, a rigorous and practical performance evaluation methodology is required to be established. In what follows, these two issues will be dealt with sequentially.

\subsection{Keyless Secure Transmission Architecture Design}
The proposed keyless secure transmission architecture is shown in Fig. 5, which consists of communications modules and security modules. The blue parts in this figure are communications modules, which implement physical layer security algorithms (such as secure beamforming, artificial noise injection, etc.) \cite{AN} and are used to provide a basic secure transmission link. The yellow parts correspond to security modules, which are realized using cryptographic primitives \cite{Shafi}. The inclusion of the security modules enables the system to achieve a provable security strength.

\begin{figure}[!t]
\centering
\includegraphics[width=5.5in]{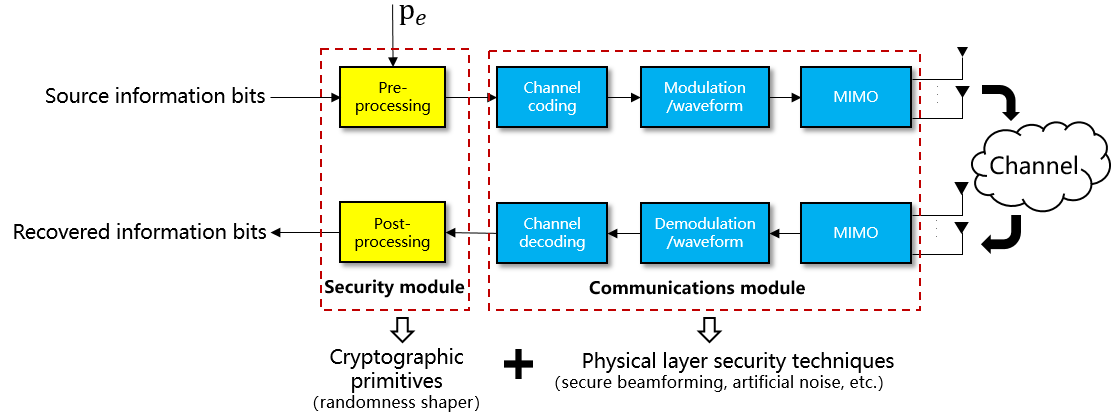}
\caption{Keyless secure transmission architecture.}
\label{fig_5}
\end{figure}

This architecture integrates cryptographic approaches (but different from classical encryption and decryption techniques) with PLS techniques to realize a keyless secure transmission mechanism. In this architecture, PLS techniques are first applied to create a high error floor at the eavesdropper, i.e., introduce some randomness into the wiretap channel. Then, the pre-processing module, which actually acts as a ``randomness shaper", extracts and diffuses the aforementioned randomness introduced by PLS algorithms. This module essentially produces a virtual key that follows a uniform distribution, which can be used to achieve a provable security level for all the transmission bits within any information block. Note that the pre-processing module has an input parameter $p_{e}$ representing the error floor created at the eavesdropper. Based on this information, the system is able to adjust its transmission parameters such as the length of a block, the details of which will be given later. In practical systems where the location of the eavesdropper is typically unknown, it is impossible to obtain the error probability of the eavesdropper. However, for any given physical layer security algorithm and channel model, we can estimate the error floors created at all possible positions of the eavesdropper, and use the minimum of all these error floor values as the aforementioned $p_{e}$.

\subsection{Enabling Technology: Randomness Shaping (RS)}
The core component in the developed keyless secure transmission architecture is the randomness shaper in the pre-processing module. One possible implementation of the randomness shaper is depicted in Fig. 6, where $m_{1}, m_{2}$, ..., $m_{q}$ stand for the information blocks to be transmitted, and $x_{1}, x_{2}$, ..., $x_{q}$ denote their corresponding coded versions, i.e., the outputs of the channel encoders. The lower half of Fig. 6 corresponds to the error control coding (ECC), where various channel codes can be adopted, such as LDPC, Polar, etc. The upper half in Fig. 6 is used for randomness shaping (RS), where ORE, BRE, and CRE are one-way randomness extractor, bidirectional randomness extractor, and compressive randomness extractor, respectively. $t_{0}$ is an initial random vector that is transmitted before the information blocks.

\begin{figure}[!t]
\centering
\includegraphics[width=5in]{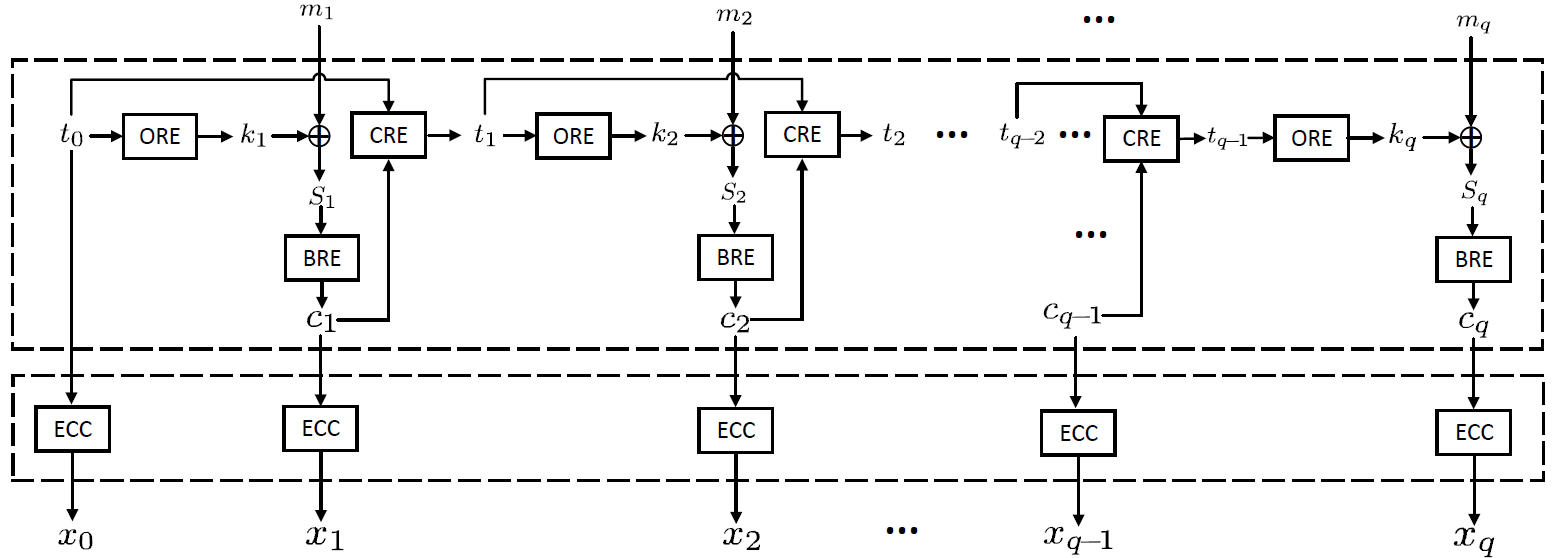}
\caption{An embodiment of the randomness shaping module.}
\label{fig_6}
\end{figure}

The principle of the randomness shaping is as follows: First, within each block, the BRE is used to extract and diffuse the randomness that is introduced in the wiretap channel by physical layer security approaches. With the help of BRE, all information bits in this block can be well protected, i.e., the decoding errors created by physical layer security methods can be distributed across the entire block uniformly. Then, CRE and ORE in each block further accumulate the decoding noises in preceding blocks and aggregate them with the decoding noises in the current block. Thanks to this ``error accumulation" mechanism, the eavesdropper cannot recover the current block even if the channel condition experienced by this block is pretty good such that PLS technique is unable to introduce enough randomness. In short, the adoption of CRE and ORE well solves the problem that the required security strength cannot be achieved for some blocks because of the lack of randomness.

Assume that the error floor introduced by physical layer security algorithms at the eavesdropper is $p_{e}$, and the security level in terms of computational complexity is $\lambda$ (i.e., $2^{\lambda}$ tries are required for the attacker to decipher the block). Then, for the randomness shaper exhibited in Fig. 6, the block length \emph{L} should satisfy:
\begin{equation}
\label{eq-5}
L\ge\frac{\lambda}{\log \frac{1}{1-p_{e}}},
\end{equation}
from which we know that the choice of the block length relates to the error floor and the target security level. This provides useful insights into the communications system design from a security perspective. That is, the transmitter can realize flexible security levels by adjusting the block length.

\subsection{Performance Evaluation}
In this subsection, the performance of the keyless secure transmission architecture will be evaluated theoretically using the proposed DoA and DoSA metrics. Because explicit key generation procedure is not needed, it is apparent that DoA is equal to DoSA for the keyless architecture. So, only DoSA will be analyzed in the following discussions.

Suppose the length of each block is \emph{L} bits, and for every received bit, the min-entropy at the eavesdropper caused by physical layer decoding errors is $H_{\infty}(X|Z)$. Then, the introduced min-entropy at the eavesdropper for the entire block is $H_{\infty}(X|Z) \times L$ bits, which is actually the length of the virtual key. By definition, the achievable DoSA is expressed as
\begin{equation}
\label{eq-6}
d=\frac{H_{\infty}(X|Z) \times L}{L}= H_{\infty}(X|Z),
\end{equation}
which can be further upper bounded by
\begin{equation}
\label{eq-7}
H_{\infty}(X|Z)\le \log\frac{1}{1-p_{e}}.
\end{equation}
Here, $p_e$ is the error floor at the eavesdropper that is created by physical layer security techniques. Based on (7), we can relate the security strength to the error floor at the eavesdropper. More specifically, by combining (6) and (7), we have
\begin{equation}
\label{eq-8}
p_{e}\ge 1-2^{-d},
\end{equation}
which indicates that if the system's requirement on security is $d\ge d_{0}$, the introduced error floor by physical layer security algorithms at the eavesdropper should be at least $1-2^{-d_{0}}$. Based on this, specific physical layer security algorithms can be developed.

In general, only an upper bound for the achievable DoSA can be obtained. However, given the structure of randomness shapper, the exact DoSA may be derived. For example, the DoSA for the randomness shapper displayed in Fig. 7 is $\frac{\lambda}{L}$. Assuming $\lambda=128$, we have:

\begin{itemize}
  \item If $p_{e}=0.2$, \emph{L} should be at least 398, yielding the DoSA to be 0.32;
  \item If $p_{e}=0.3$, \emph{L} should be at least 249, yielding the DoSA to be 0.51;
  \item If $p_{e}=0.5$, \emph{L} should be at least 128, yielding the DoSA to be 1, i.e., one-time pad perfect secrecy.
\end{itemize}

The above analysis clearly showcase that by using the proposed keyless secure transmission architecture, we can establish the relationship between the error floor at the eavesdropper and the DoSA, which relates the communications metric (e.g., BER) to security strength. This answers a fundamental question which is open for several decades: How to evaluate the security performance of physical layer security techniques both rigorously and practically?

Some preliminary simulations are carried out to verify the effectiveness of the proposed keyless secure transmission method. In the simulations, a typical three-node wiretap channel is considered, and $\lambda$ is set to be 128. Both the legitimate and the eavesdropping links are modeled as binary symmetric channels (BSC). The crossover probability of the legitimate channel is $10^{-3}$, and that of the eavesdropping channel varies from 0.2 to 0.5. Here, the ``physical advantage" of the legitimate channel can be easily obtained using the existing PLS techniques, e.g., the artificial noise injection method proposed in \cite{AN}. Numerical results are presented in Table II, where ``XX\_WO" and ``XX\_W" denote the achievable performance without randomness shaper and that using randomness shaper, respectively. It is observed that Eve's BER can be increased to 0.5 by combining PLS methods with randomness shaping, which demonstrates that the inclusion of randomness shaping enhances transmission security significantly.

\begin{table}[ht]
\begin{center}
\caption{\label{tab:table1}Simulation results of the keyless secure transmission scheme}
\begin{tabular}{cccc}
\toprule
BER\_Eve\_WO & BER\_Eve\_W & Block length \\
\midrule
0.2 & 0.5 & 398\\
0.3 & 0.5 & 249\\
0.4 & 0.5 & 174\\
0.5 & 0.5 & 128\\
\bottomrule
\end{tabular}
\end{center}
\end{table}

\section{Comparisons of Different Secure Transmission Solutions}
In this section, we compare different secure transmission solutions (i.e., OTP, traditional cryptographic approaches, PLS, and the proposed Approaching-OTP methods) from four perspectives, as is shown in Table III. In addition to DoA and DoSA that are explained previously and Implementation Complexity that is easily understandable, Semantic Security (SS) \cite{Shafi}, which is a widely-accepted metric in security community, is also adopted here as a performance indicator.

The main points of Table III are summarized below. (1) OTP achieves highest security level at the cost of an unaffordable complexity; (2) Classical cryptographic approaches are easily implementable. However, there is a huge gap between its achievable security strength and perfect secrecy; (3) Despite its relatively-low implementation complexity, PLS cannot obtain provable security performance or guarantee semantic security; (4) The proposed Approaching-OTP methods (NBKG, RS+PLS), for which the achievable security performance can be assessed quantitatively, well balance performance and cost.

\begin{table}[ht]
\begin{center}
\caption{\label{tab:table1}Comparisons of different secure transmission solutions}
\begin{tabular}{ccccc}
\toprule
  & DoA & DoSA & Complexity & SS \\
\midrule
OTP & 1 & / & High & Yes\\
NBKG & close to 1 & see E.q.(4) & Medium & Yes\\
RS+PLS & see E.q.(7) & see E.q.(7) & Medium & Yes \\
Cryptography & $ \ll 1$ & / & Low & Yes\\
PLS & / & / & Low & No\\
\bottomrule
\end{tabular}
\end{center}
\end{table}

\section{Concluding Remarks}
In this article, the topic of approaching one-time pad perfect secrecy is discussed. While one-time pad, originally proposed by Shannon, has appeared for more than 70 years, little work has been done to investigate how to efficiently ``approach" it. To deal with this fundamental problem, we first present the definition of ``approaching one-time pad", for which two performance metrics, namely Degree-of-Approaching and Degree-of-Synchronous-Approaching, are developed to respectively characterize the closeness to perfect secrecy and the system's ability in approaching it. Afterwards, we propose two paths towards one-time pad, i.e., the explicit encryption based method and the implicit encryption based method. For the former, we are mainly concerned about high-rate key generation, for which a novel framework called NBKG is developed. For the latter, we devise a keyless secure transmission architecture, where physical layer security schemes are combined with a so-called randomness shaping technique to provide a provable security level. The concepts and techniques developed in this article shed new lights into the wireless security research, and provide promising solutions for post-quantum security.

\end{document}